# QoE Enhancing Schemes for Video in Converged OFDMA Wireless Networks and EPON

Divya Chitimalla[1], Massimo Tornatore[1, 2], Sang-Soo Lee[3], Han-Hyub Lee[3], Soomyung Park[3], HwanSeok Chung[3], and Biswanath Mukherjee[1]
[1]University of California Davis, USA; [2]Politecnico di Milano, Italy;
[3]Electronics and Telecommunications Research Institute, Korea;
{dchitimalla, mtornatore, bmukherjee}@ucdavis.edu,
{soolee, hanhyub, smpahk, chung}@etri.re.kr

*Abstract*—Bandwidth requirements of both wireless and wired clients in access networks continue to increase rapidly, primarily due to the growth of video traffic. Application awareness can be utilized in access networks to optimize Quality of Experience (QoE) of end clients. In this study, we utilize information at the client side application (e.g., video resolution) to achieve superior resource allocation that improves users' QoE. We emphasize on optimizing QoE of the system rather than Quality of Service (QoS) as user satisfaction directly relies on QoE, and optimizing QoS does not necessarily optimize QoE as shown in this study. We propose application-aware resource allocation schemes on an Ethernet Passive Optical Network (EPON), which supports wireless (utilizing Orthogonal Frequency Division Multiple Access (OFDMA)) and wired clients running video-conference applications. Numerical results show that the application-aware resource allocation schemes improve QoE for video-conference applications for wired and wireless clients.

## I. INTRODUCTION

Communication networks connect an enormous number of devices (smartphones, tablet, etc.) requesting continuously-increasing bandwidth. Some of the major applications today are video-based, such as Video on Demand (VoD), live video streaming, telemedicine, security monitoring, etc. These video-based high-bandwidth applications take up to 55% bandwidth of current wireless networks (e.g., 4G Long-Term Evolution (LTE)) [1] and are expected to consume up to 75% bandwidth in future wireless networks (5G) [2].

Unfortunately, this growth of bandwidth demand does not match with an increase in available access capacity, which makes access bandwidth a bottleneck. Cisco Global Mobile Data Traffic Forecast Update (2016–2021) predicts monthly global mobile data traffic will be 49 exabytes by 2021, leading to an annual traffic that will exceed half a zettabyte. This is a challenging target for mobile network operators that are currently investigating new network solutions to address such traffic growth. For example, the cloud-RAN architecture is being investigated by many leading telecom players [3] [4] [5], confirming that access bandwidth is becoming a constrained resource given the huge traffic growth. Network operators need resource-optimization techniques that improve service delivery to end clients while keeping their expenses low. Hence, effective mechanisms for allocation of access bandwidth (both wireless and wired) are sought such that video clients get good services with high quality. We address the access bottleneck issue by evolving the network from being application-agnostic to application-aware, to provide better Quality of Experience (QoE) for video clients. Video clients are classified into different types based on their Service Level Agreement (SLA) as discussed in Section III. We note that QoE optimization is highly dependent on the application characteristics (such as video-streaming procedures). Quality of Service (QoS) does not necessarily capture user satisfaction as there is no linear relation between QoS and QoE, as discussed in this study.

### A. Application-Aware Networking for QoE Optimization in Video Clients

An application-aware network keeps track of important information about applications that are running on it, to optimize its performance [6]. This is a promising solution to achieve superior QoE and bandwidth utilization. It can benefit from Software-Defined Networking (SDN), which enables global management of a network by abstracting its lower-level functions [7] and decoupling the control plane from data plane. In this way, an application-aware network can directly correlate the resource-allocation strategies in the access network with QoE of clients.

Quality of Service (QoS) parameters such as delay, jitter, packet loss, etc. cannot exactly quantify the user experience as there is no linear relation between QoS and QoE [8]. We quantify QoE using parameters such as empirical Mean Opinion Score (MoS) and probability of call drop due to user unsatisfaction.

Video quality is divided into different profiles associated to different video resolution and frames per second (FPS), and there is no linear relation between MoS and bandwidth for video applications. Each video-quality profile is mapped over a specific bandwidth range, and any bandwidth that falls in that range maps to the same QoE. Hence, using the lowest bandwidth value in this range, we can achieve same QoE as the highest bandwidth in this range.

To identify the bandwidth-range profiles associated to each QoE level, we need to consider how video-streaming bandwidth is adaptively assigned in a specific video application, e.g., Skype. Once the (non-linear) relation between QoE and bandwidth for a specific application is known, our application-aware approach can make intelligent decisions on bandwidth allocation to the users to maximize QoE. In summary, even though QoE is dependent on QoS, since they are not linearly related, we cannot perform resource assignment that maximizes QoE for all users, if we just focus on QoS. Note that QoS maximization could be done without application awareness, whereas QoE maximization requires application awareness.

### B. Wired–Wireless Converged Optical Access Network Architecture

Optical access network such as EPON is a dominant technology, thanks to its low capital (CAPEX) and operational (OPEX) expenditures [9]. It is traditionally employed to support wired residential and business users (fiber-to-the-x (FTTx) clients, where x can be H (home), B (business), etc.). It consists of an Optical Line Terminal (OLT) connected to Optical Network Units (ONUs) through a feeder fiber, a splitter, and distribution fibers. In EPON, the downstream traffic going from the OLT to the ONUs is broadcast, while the upstream traffic is multiplexed in the feeder fiber using a dynamic medium access control (MAC) protocol, such as Interleaved Polling with Adaptive Cycle Time (IPACT) [10]. EPON is a strong candidate to provide Radio Access Network (RAN) infrastructure to both legacy wireless technologies [11] (4G LTE or 3G) and future wireless technologies (5G and beyond) [12]. EPON can be used to backhaul traffic from small cells and macro cells to the mobile-core network due to its support for high bit rates and flexible bandwidth allocation [13].

The wired-wireless converged EPON architecture considered in this study is shown in Fig. 1. This architecture can be beneficial in situations where base stations in business areas (e.g., crowded high-speed rail regions, shopping malls, etc.) are supported along with residential wired users. Typically, wireless networks in business areas are heavily loaded in the morning, whereas residential-area wired networks get most of their traffic in the evening, leading to "tidal effect" of the network traffic. Joint resource allocation in the wired-wireless converged EPON can benefit from such "tidal effect".

### C. Contribution of This Study

The contribution of this study is "two-fold" in the sense that our proposed QoE aware bandwidth allocation has two dimensions of resource assignment: wireless resources and EPON backhaul bandwidth. First, we propose application-aware wireless resource block allocation schemes (using orthogonal frequency division multiple access (OFDMA)) that maximize the cumulative QoE (using MoS) of wireless clients after guaranteeing a minimum QoE to different client types to satisfy the user SLA as discussed in Section III. Second, we study application-aware wired-wireless converged EPON that provides superior QoE for each client type (see Table 2), along with dynamic backhaul bandwidth depending on wireless allocation to wireless clients. We exploit "tidal effect" of wired and wireless networks supported by EPON, where a converged architecture would be beneficial. Numerical results show improvement of the empirical MoS. Without loss of generality, our work considers Skype (video-conference application) characteristics to perform application-aware resource allocation, but this can be extended to other applications (e.g., video streaming).

The rest of this study is organized as follows. Section II gives an overview of application-aware wired-wireless converged EPON. Section III outlines a video-conference application such as Skype, which is reverse engineered [8] to provide an empirical relation between application parameters such as video resolution, FPS, video rate, and empirical QoE measure, i.e., MoS. Section IV-A studies application-aware wireless resource-allocation strategies that enhance QoE of wireless clients. Section IV-B outlines QoE enhancing resource allocation schemes at the OLT utilizing application information to allocate bandwidth jointly for wireless backhaul and wired clients. Section V provides the numerical evaluation of the proposed schemes for both wireless and wired parts. Results show that the proposed schemes in application-aware converged EPON outperform application-unaware schemes. Section VI concludes the study.

There are several studies that provide solutions for resource allocation in hybrid PON-LTE architectures. Ref. [14] provides a survey on dynamic bandwidth allocation (DBA) schemes in optical networks that support QoS optimized scheduling in wireless networks. Ref. [15] utilizes XG-PON standard-compliant bandwidth to ensure QoS for combined voice and video traffic. Ref. [16] provides a hierarchical QoS-aware dynamic bandwidth allocation algorithm which supports bandwidth fairness at the ONU base station level and distinguished user QoS requirements for different traffics. Ref. [17] explores PON based RAN architecture for LTE mobile backhaul networks employing ring-based WDM PONs which monitors the traffic imbalances of downstream channels for load balancing, by dynamically reallocating and sharing the capacity of the downstream wavelength. However, there are no studies which employ application-awareness to closely monitor video traffic and perform resource allocation that maximizes QoE for converged PON-OFDMA networks.

## II. APPLICATION-AWARE EPON WIRED-WIRELESS CONVERGED ACCESS NETWORK

Figure 2 shows the application-aware wired-wireless converged EPON architecture considered in this work. This architecture can support small cells as well as regular eNodeBs, by backhauling the traffic over EPON. Here, the traditional OLT-ONU framework is unaltered, but there is a feedback mechanism from wired clients to a fixed SDN controller, and from wireless clients to a mobile SDN controller, interfacing with the wireless Medium Access Control (MAC) resource scheduler, co-located with the OLT as shown in Fig. 2.

Our study utilizes application information to make intelligent decisions on resource allocation that maximize QoE. Control signals are routed from clients to the mobile and fixed SDN controllers using uplink bandwidth for both wired and wireless networks. Figure 2 also shows the separated data plane, control plane, and service plane of the proposed architecture. An SDN management application collects feedback from client applications (as discussed in our previous work [18] [19]) and feeds them to the SDN controller through a North-Bound API, which is the interface between the control and application layers. The fixed and mobile SDN controllers can interface with both the OLT and wireless MAC scheduler (for resource block allocation in OFDMA) by a South-Bound API.

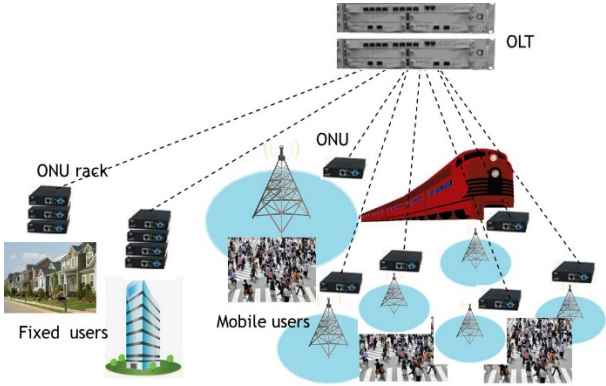

Figure 1: Converged access network supporting wired and wireless clients.

The links (colored lines) from ONU to clients can be optical fiber, copper, or wireless (for cellular transmissions). The links (dotted lines) between wired clients, fixed SDN controller, and OLT are logical connections for sending application feedback from wired clients; and the links from fixed controller to OLT are for controlling the OLT's scheduling. The links (dotted lines for wireless base station) between wireless clients and mobile SDN controller are logical connections for sending wireless application feedback. The overhead to send application information for wireless is small, as wireless users exchange information with the scheduler such as channel quality, and the application information can be piggybacked as an additional field that captures QoE of the user.

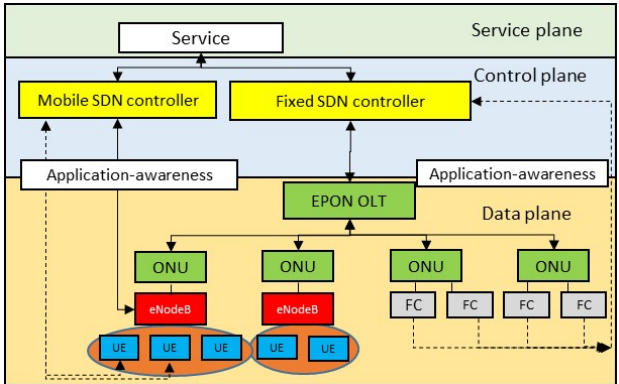

Figure 2: Application-aware SDN-enabled wired-wireless converged EPON.

At the SDN controller, a table is maintained with client application information and is updated every time it receives a new control packet. The table has the current FPS and resolution information for all video clients. Based on this information, the controller evaluates the video rate (i.e., bit rate at which video is sent out) and hence sending rate (bit rate of video along with redundancy bits to protect against packet loss) along with MoS, as shown in Table 2.

The OLT performs downstream bandwidth allocation by using Time-Division Multiplexing (TDM), and most popular methods include round robin, weighted round robin, and excess redistribution [20] [21] [22] [23]. To the best of our knowledge, existing proposals for resource allocation in EPONs have not considered service differentiation based on QoE of clients using application awareness. The MAC layer of an LTE scheduler allocates resources on a Transmission Time Interval (TTI) timescale, taking into account items such as delay tolerance, error resilience, and bit rate, via a parameter called QoS Class Indicator (QCI). Typical LTE scheduling mechanisms include round robin, proportionally fair, greedy maximization, etc. [24].

Table 1: Notations.

| Video rate | $R_V$ |
|---|---|
| Resolution | R |
| Frames per second | FPS |
| Sending rate | $R_S$ |
| FEC redundancy factor | F |
| Quantization parameter | Q |

However, these schemes are application agnostic, and the network decides the resource allocation based on QoS parameters and does not use QoE evaluation. Such an application-agnostic resource allocation gives sub-optimal user experience, when bandwidth becomes constrained. Hence, there is a need for superior resource-allocation schemes that use application awareness to maximize QoE of clients as QoS optimization may not directly result in optimal QoE as discussed in Section I.

### III. EMPIRICAL MOS AND CLIENT CATEGORIZATION

Empirical MoS is a measure of QoE of clients (Skype's subjective quality specified by ITU-T Recommendation G.1070 [8]), and it can be calculated using video rate ($R_v$) and FPS as shown below:

$$Q_s = 1 + \left(c - \frac{c}{\left(1 + \frac{R_v}{d}\right)^e}\right) * exp\left(-\frac{(ln(FPS) - ln(a + b * R_v))^2}{(2 * h + g * R_v)^2}\right) \quad (1)$$

where a = 1.43, b = 0.02, c = 3.75, d = 184.1, e = 1.16, h = 1.44, and g = 0.0388.

Table 1 describes the notations utilized in this section. A Skype profile is characterized by its resolution, FPS, and video rate. Each profile corresponds to an empirical MoS value. Table 2 gives resolution, FPS, video rate, sending rate, MoS, and call drop due to user unsatisfaction, for different Skype profiles. These values are obtained from [8], which is a thorough experimental analysis on Skype video measurement.

Ref. [8] studies the user back-off probability (call drop due to unsatisfaction) which is the probability that the user turns off his/her video call due to bad Skype quality. Note that this probability is different than blocking probability due to resource exhaustion as, e.g., a call session might still continue with just voice after video has been dropped). Call drop due to user unsatisfaction is probabilistically measured with MoS, and it turns out to be a function of the MoS as shown in Table 2. This probability is specific to Skype application and not only depends on QoS parameters but also on the application design (e.g., Skype's adaptive video streaming algorithm), and is a subjective measure to QoE as studied for Skype calls. MoS ranges from 2.9 to 4.7 (with increasing QoE), where MoS of 2.9 indicates poor QoE, and MoS of 4.7 indicates best QoE [8]. MoS greater than 4.5 yields 0% call drop due to user unsatisfaction.

Skype uses multi-rate encoding and adapts to network conditions (as in DASH (Dynamic Adaptive Streaming over HTTP) [24]). We assume that Skype video adapts to the profile from Table 2, which has the best possible FPS, and resolution whose bandwidth requirement is lesser than the available network bandwidth [8]. Given FPS and resolution, video rate is calculated as follows, where quantization parameter is set to 2 bytes per pixel:

$$R_v = R * FPS * Q \quad (2)$$

For a packet loss of 5%, $FEC\ redundancy\ factor$ is 40% for Skype[8]. Send rate ($R_s$) is calculated from video rate using:

$$R_S = R_v/(1 - FEC) \quad (3)$$

Table 2: Resolution, frame rate, video rate, MoS, and call drop for Skype application profiles [8].

| Resolution (R) | FPS | Video Rate ($R_v$) | Send Rate ($R_S$) | MoS | Call Drop |
|---|---|---|---|---|---|
| 640*480 | 5 | 3.07 | 5.12 | 4.62 | 0.000 |
|  | 10 | 6.14 | 10.23 | 4.69 | 0.000 |
|  | 15 | 9.20 | 15.33 | 4.72 | 0.000 |
|  | 28 | 17.20 | 28.66 | 4.74 | 0.000 |
| 320*240 | 5 | 0.77 | 1.28 | 4.15 | 0.070 |
|  | 10 | 1.54 | 2.56 | 4.46 | 0.008 |
|  | 15 | 2.30 | 3.83 | 4.57 | 0.000 |
|  | 28 | 4.30 | 7.16 | 4.66 | 0.000 |
| 160*120 | 5 | 0.19 | 0.32 | 2.92 | 0.532 |
|  | 10 | 0.38 | 0.64 | 3.63 | 0.248 |
|  | 15 | 0.57 | 0.96 | 3.96 | 0.116 |
|  | 28 | 1.07 | 1.70 | 4.32 | 0.036 |

Ref. [25] presents a measurement study on three-popular video-telephony applications: Skype, Google+, and iChat. All these applications adapt traffic to network conditions using adaptive video streaming and by storing video chunks encoded into multiple layers as in Scalable Video Coding (SVC). To allow for streaming, an SVC stream is divided into chunks. Each chunk contains layers in the three-dimensional quality space including temporal, spatial and, SNR scalability [26].

However, these applications differ from each other in terms of video resolutions and supported frame rate (see Table 5). The exact algorithms employed are confidential, however Ref [25] provides useful metrics that can be utilized in this study to perform QoE-aware resource allocation.

Ref. [25] shows also that the frame rates can vary from 1 FPS to 30 FPS for each system, and observed resolution values are also listed in Table 5. Skype and Google+ adapt video resolution to network bandwidth. iChat's video resolution is determined by the number of users in the conference. For example, the resolution is always set to be 640 × 480 in the case of a two-party call, and once the resolution is set at the beginning, it will not be changed afterward, no matter how we change the bandwidth setting.

To provide better services to clients, the network provider can consider different requirement for each client based on a Service Level Agreement (SLA). Here, we consider that a minimum MoS must be provided to each client to ensure that all clients get a satisfactory QoE, even when bad wireless channel conditions might degrade QoE for some users. Once this minimum MoS is guaranteed, our algorithm will maximize overall MoS in the system. That is, if a client has subscribed for the highest-quality video service for an important business conference, he/she must be assured the best possible QoE (high MoS), and be given differentiated treatment from other clients who can tolerate lower QoE for video (e.g., users having Skype for free).

Table 3: Client types and acceptable MoS ranges.

| Client Type | Acceptable MoS |
|---|---|
| 3 | 4.1-4.7 |
| 2 | 3.5-4.7 |
| 1 | 2.9-4.7 |

Table 4: Example of mapping between resource blocks and MoS for three client types with different channel.

| Client type 1 | | Client type 2 | | Client type 3 | |
|---|---|---|---|---|---|
| #RBs | MoS | #RBs | MoS | #RBs | MoS |
| 1 | 2.9251 | 1 | 3.5376 | 1 | 4.1269 |
| 1 | 3.4222 | 1 | 3.7955 | 1 | 4.1856 |
| 1 | 3.8606 | 1 | 4.0560 | 2 | 4.2772 |
| 1 | 4.0932 | 2 | 4.1856 | 2 | 4.4280 |
| 1 | 4.2114 | 2 | 4.2772 | 3 | 4.5538 |
| 2 | 4.2960 | 2 | 4.4280 | 4 | 4.6127 |
| 2 | 4.4632 | 3 | 4.5538 | 4 | 4.6409 |
| 3 | 4.5688 | 4 | 4.6127 | 5 | 4.6605 |
| 4 | 4.6209 | 5 | 4.6409 | 8 | 4.7024 |
| 5 | 4.6465 | 6 | 4.6605 | 11 | 4.7222 |
| 5 | 4.6641 | 10 | 4.7024 | 15 | 4.7321 |
| 9 | 4.7077 | 14 | 4.7222 | 18 | 4.7380 |

Table 5. Supported video rates and resolutions for iChat, Google+, and Skype [25].

| Application | Rate (kbps) | Resolutions |
|---|---|---|
| iChat | 49 – 753 | 640*480,320*240,160*120 |
| Google+ | 28 - 890 | 640*360,480*270,320*180 240*135,160*90,80*44 |
| Skype | 5 – 1200 | 640*480,320*240,160*120 |

To illustrate this client-level differentiation, we have classified the clients into three categories based on the level of QoE supported by network. A user of Category 3 is the highest-priority user, who accepts only very high MoS. The MoS range is split into three parts. A client of Type 1 accepts any MoS ranging from 2.9 to 4.7. Hence, he/she pays lesser for the service compared to a client of Type 2, since the latter's minimum acceptable MoS is higher (for Type 2, minimum MoS is 3.5, i.e., range of 3.5 to 4.7, and for Type 3 it is 4.1, i.e., range of 4.1 to 4.7). Hence, different revenues can be collected from clients of different types to achieve client-level service differentiation. Revenue maximization for such a scenario is an open problem for future research. The relation between user types and acceptable MoS is summarized in Table 3. Note that the same concept can be extended to other applications, which do not have stringent bandwidth requirements.

## IV. RESOURCE ALLOCATION STRATEGIES

Resource allocation in converged OFDMA wireless and EPON is a two-step process. In Section IV-A, we provide application-aware OFDMA resource-allocation algorithms, which aim to maximize QoE of wireless clients. In Section

IV-B, we provide application-aware EPON resource-allocation algorithms that support both wireless and wired clients. We assume that EPON allocates enough backhaul bandwidth to support wireless bit rates as determined by wireless resource-allocation strategies in Section IV-A and strives to maximize QoE of wired clients using the remaining EPON bandwidth.

### A. Application-Aware Wireless (OFDMA) Resource Allocation for QoE optimization

Wireless networks (4G and anticipated in 5G) utilize OFDMA where each LTE downlink frame has a duration of 10 milliseconds (ms). Each frame is further subdivided into ten 1-ms Transmission Time Intervals (TTIs). The minimum unit of resource assignment for user equipment (UE) is one Physical Resource Block (PRB), consisting of one TTI and one sub-channel of bandwidth of 180 kHz, thus enabling fine-grained scheduling and radio resource control.

Total number of resource blocks in the downlink depends on the wireless spectrum bandwidth, as specified in 3GPP [27]. 20 MHz wireless spectrum is considered in this study, which corresponds to 100 PRBs. Maximum data rate (on physical layer) supported by the channel is decided based on channel quality indicator (CQI) which varies from 0-26 where CQI of 0 represents a poor channel and 26 represents a superior channel. A better channel (and hence higher CQI) supports higher data rates (using higher modulation format or transport block size (TBS)). We considered Rayleigh fading channel along with path loss in the model. The signal is generated and passed through a Rayleigh fading channel. The output signal is discretized to reflect CQI value from 0 to 15. Based on CQI, transport block size (TBS) and hence the coding rate are calculated as in [23]. Signal strength at the receiver and throughput, i.e., coding rate, depend on channel condition; hence, QoE of video is hugely affected by the channel. A bad channel would result in low CQI, low TBS, and low throughput, which mean more resource blocks are required for the UE to get acceptable QoE.

Each Skype profile can be supported by giving a sufficient number of PRBs on the downlink, based on the channel condition. The wireless channel condition varies rapidly due to multipath fading considered in our evaluation system. A poor channel requires more PRBs to achieve the same video quality compared to a high-quality channel. There is a need for intelligent resource allocation considering channel qualities, client types, and MoS to maximize overall QoE (cumulative MoS, i.e., MoS for each client added up) of the system while maintaining a minimum acceptable MoS at each client to satisfy client SLA. This requires continuous monitoring of the applications and carefully balancing the PRB allocations.

### A.1 Wireless Resource Allocation Strategies

In this subsection, some parameters and concepts of wireless resource-allocation schemes are introduced. The total number of resource blocks for a given wireless spectrum is fixed which is also known as the budget of the system, and is denoted by T. Total number of resource blocks currently allocated to the clients, utilizing any strategy, is denoted by B. Each client is denoted by a type 'i', where each type corresponds to a minimum acceptable MoS. In each PRB allocation cycle, the channel condition (CQI) at each UE is reported to the cell, and the supported TBS is assigned to the UE (as given by the 3GPP TBS mapping).

A mapping is formed between the number of PRBs allocated to the UE (an example is shown in Table 4) and the MoS that can be supported (with the corresponding data rate). The number of resource blocks needed to achieve a certain Skype profile is known as weight, and the MoS supported is known as profit. There is a one-to-one mapping between number of resource blocks (if we remove the physical-layer redundancies), i.e., weights, and MoS supported, i.e., profits. We also consider the minimum MoS acceptable at each client's SLA to form this weight-profit mapping. Below, we propose three algorithms that consider the application characteristics to achieve optimal QoE.

### A.2 Modified Round Robin (MRR)

The mapping between weights and profits is formed as described in Section IV-A.1 for each client type. In MRR, we first allocate a minimum number of PRBs so that minimum acceptable MoS (first entry of Table 4) is achieved at all clients (step 1). We choose a random UE using a uniform distribution, and assign the best profile within un-allocated budget that maximizes its profit. This process is continued in a round-robin fashion until un-allocated bandwidth (T-B) becomes zero or all the UEs get maximum possible MoS.

### A.3 Water-Filling Algorithm (WF)

UEs are sorted in decreasing order of channel conditions (transport block sizes, and hence maximum transmission rate supported). We choose the first UE (which has the best channel condition) and allocate the best profile within un-allocated budget that maximizes MoS of the UE. We proceed to the next UE (next best channel) and repeat this process until all resources are allocated or all UEs have their highest MoS levels. This algorithm utilizes channel information to achieve high data rates (and hence high MoS) on the UEs which have superior channel conditions. This algorithm is called Water-Filling as the UE with best channel gets best MoS just like water that flows and settles at the lowest surface. It is similar to greedy maximization in regular MAC scheduling algorithms, however the resource allocation now utilizes application information.

---

**Algorithm 1: Modified Round Robin**
**Input:** client type $c_i$, channel condition $cqi_i$, total PRBs T
**Output**: PRB assignment for each UE, and cumulative MoS
  *step 1:* Allocate resources such that minimum acceptable profiles are supported for all UE
  *step 2:* Calculate allocated budget in step 1, call it B
  *step 3:* Choose a random UE number using uniform distribution
  *step 4:* Allocate best profile within remaining budget (T-B) that increases MoS of UE
  *step 5:* Repeat step 4 in round robin fashion until all resources are allocated or all UEs are visited

**Algorithm 2: Water-Filling**
**Input:** client type $c_i$, channel condition $cqi_i$, total PRBs T
**Output:** PRB assignment for each UE, and cumulative MoS
  *step 1:* Allocate resources such that minimum acceptable profiles are supported for all UE
  *step 2:* Calculate allocated budget in step 1, call it B
  *step 3:* Sort UEs in decreasing channel conditions (coding rate)
  *step 4:* Select the UE with maximum coding rate
  *step 5:* Allocate best profile within remaining budget (T-B) that increases MoS of UE
  *step 6:* Repeat step 4 in descending order of channel rates until all resources are allocated or all UEs are visited

**Algorithm 3: Multiple-Choice Knapsack Algorithm**
**Input:** Client type $c_i$, channel condition $cqi_i$, total PRBs T
**Output:** PRB assignment for each UE, and cumulative MoS
  *step 1:* (a) For every UE, determine weights and profits based on their client type and channel condition
  (b) Weight is number of resource blocks needed to achieve a certain video rate and hence certain MoS
  (c) Profit is the incremental MoS achieved by using certain profile level; scale profits based on scaling factor of the algorithm and round it off to near integer
  *step 2:* For every UE, determine channel condition (every TTI), based on which maximum coding rate is calculated
  *step 3:* Using this coding rate, find the acceptable weights and profits for each UE
  *step 4:* (a) For every UE, pick one profile level (w,p) such that cumulative MoS is maximized using dynamic programming
  (b) Compute upper bound for cumulative MoS, call it Q
  (c) Set $Y_0(0) = 0, Y_0(q) = T+1$ for all q from 1 to Q,
    **for** i = 0 to N
      **for** q = 1 to U
        $Y_i(q) = min \{ Y_{i-1}(q - p_{ij}) + w_{ij} \}$
      **end**
    **end**

---

### A.4 Multiple-Choice Knapsack Algorithm (MCKP)

The resource allocation, where the optimization objective is to achieve maximal cumulative MoS (hence QoE) while maintaining the minimal acceptable MoS for each UE (according to SLA), can be proved to be NP complete. This optimization can be modelled as a multiple-choice knapsack problem, where there are a fixed number of buckets, and each bucket has a finite number of objects having weights/value pairs. The idea is to pick exactly one object from each bucket such that overall value is maximized while not exceeding total budget (weight). We denote weights and profits in the same way as in Section IV-A.1. We use dynamic programming with scaling profits [28] to solve the multiple-choice knapsack problem as described below.

$Y_i(q)$ is defined as minimum number of resource blocks required to achieve a cumulative MoS of $q$ using UEs from 1 to $i$. This problem can be solved by the recursive formula:
$$Y_i(q) = min\{ Y_{i-1}(q - p_{ij}) + w_{ij} \}$$
of dynamic programming, i.e., minimum resource blocks to achieve MoS of $q - p_{ij}$ for stage $i$-1 added to weight $w_{ij}$ to achieve $p_{ij}$ profit at stage $i$. Here, $j$ is the profile selected for UE '$i$'. $p_{ij}$ is profit and $w_{ij}$ is weight associated respectively for UE $i$ with profile $j$. The table entries for $Y_i(q)$ are filled for all the possible values $i$ (i.e., for all UEs in the system) and all values of $q$ (from 0 to maximum cumulative MoS achievable in the system). The maximum cumulative MoS achievable is denoted by U (which is number of UEs times the maximum MoS for each UE). Note that this optimization outputs number of resource blocks allocated to each user such that overall QoE is optimized while still providing minimum QoE following SLA for each client.

### A.5 Complexity and Optimality Analysis of Algorithms

MRR and WF are fast heuristics for application-aware wireless resource allocation. Weight vs. profit mapping is a common subroutine for all the algorithms. This mapping has worst-case time complexity of $O(T*n)$, where $T$ is total resource blocks, and $n$ is total number of wireless clients. For MRR, step 1 takes $O(n)$ to allocate minimum acceptable profile for each user. MoS enhancement steps 3, 4, 5 take worst-case complexity of $O(T*n)$ to choose the best profile achievable using remaining budget at each user. WF accounts for channel conditions to exploit the problem characteristics to enhance MoS. Step 1 of WF takes $O(T*n)$, similar to MRR. Step 3 sorts UEs in decreasing channel condition which takes $O(n*log(n))$ when merge sort or heap sort is used. MoS enhancing steps 4, 5, 6 of WF take $O(T*n)$ to allocate best profile within remaining system budget.

MRR and WF do not guarantee optimal QoE (empirical MoS) of the system. But WF should provide near-optimal results since channel condition information is exploited. MCKP solved using dynamic programming requires a table of size $n*U$ to be filled out to find the optimal cumulative MoS achievable. Worst-case complexity to determine each content of the table is $O(N*U)$, where $N$ is the sum of all possible profiles over all users. Without profit scaling, MCKP produces optimal cumulative MoS and hence optimal QoE. To make the algorithm run in a time bound by total number of users, we use profit scaling using an approximation factor €, which makes the solution within (1-€) times the optimal. Using the analysis [24] of dynamic profit scaling, we can bound the time complexity to $O(Nn^2/€)$. /€). Hence, the runtime is independent of the load in the network. Given the theoretical complexity, in our simulation, the algorithm takes less than 0.3 milliseconds to run on a computer equipped with an Intel Core i5 with processor speed: 2.4 GHz for a network with N = 427, n = 15 and $\epsilon$ = 0.1.

### B. Wired Resource Allocation for Wired-Wireless Convergence EPON Architecture

In the considered architecture (Fig. 2), we assume that each wired client has a software extension (plug-in application) that collects statistics from video-conferencing application, and reports it to fixed SDN controller, using a control packet over a control channel, as shown in Fig. 2. At the SDN controller, a table is updated with this information. The proposed EPON resource-allocation strategies assign more bandwidth to ONUs that have clients having poor QoE.

#### B.1 EPON Resource Allocation Strategies

In EPON, we utilize application-aware resource-allocation algorithm [18] to allocate resources to wired clients such that QoE is improved, after necessary bandwidth is allocated to backhaul traffic from wireless base stations. Resource-allocation strategies for wired and wireless networks are different due to inherent differences between wireless and optical channels. Wireless channel is dynamically varying due to fading and multipath propagation, whereas optical channel does not vary with time. Also, the resource allocation for wired clients cannot be done on a client-by-client basis as there is no control on individual links to the clients from OLT in EPON unlike wireless clients. So, we use average QoE of the ONU's wired clients to perform resource allocation for wired clients.

#### B.2 EPON Application-Aware Resource Allocation (EARA) algorithm

Each ONU is assigned minimum weights (and hence guaranteed minimum bandwidth) based on the number of clients and client types it supports. Based on the application information sent to the controller, empirical MOS and call-drop probability is calculated and averaged for each ONU. Excess bandwidth is the difference between total bandwidth and sum total of minimum bandwidth that must be given to each ONU and wireless backhaul resources allocated to BS. This excess bandwidth is redistributed in proportion to average probability of call drop of clients at a specific ONU.

#### B.3 Highest-Priority-First (HPF) Algorithm

HPF is an application-agnostic algorithm that gives highest bandwidth required for higher-priority clients unlike EARA which tries to maximize MoS of the system after allocating bandwidth to satisfy minimum MoS requirement of each client. Priority of an ONU is determined by considering the client types it supports. Excess bandwidth is allocated in decreasing order of priority of the ONU to attain maximum achievable MoS. This procedure is continued until all the excess bandwidth gets utilized or all clients get maximum achievable MoS.

### V. ILLUSTRATIVE NUMERICAL EXAMPLES

To evaluate the proposed application-aware resource-allocation schemes, numerical simulations are performed. We consider scenarios where five wireless bases stations (co-located with five ONUs), each with 20 MHz spectrum, along with three other ONUs supporting wired clients are connected to an OLT in an EPON.

#### A. Evaluation of Wireless Resource-Allocation Schemes

The cumulative MOS for each of our wireless resource-allocation algorithms is shown in Fig. 3. MCKP achieves the best cumulative MoS for wireless UEs. WF performs more consistently and better than MRR as it also considers channel conditions unlike MRR which randomly assigns resource blocks to UEs. MCKP can further improve MoS by compromising speed as discussed in Section IV-A.5. MoS values were rounded off to 3 significant digits to improve

---

**Algorithm 4: EPON Application-Aware Resource Allocation (EARA)**
**Input:** Data rates for each BS, total EPON bandwidth E
**Output:** Bandwidth assignment for each ONU, MoS for wired clients
    *step 1:* Allocate backhaul resources to all wireless BS based on wireless resource allocation
    *step 2:* Calculate allocated budget in step 1, call it $W_B$
    *step 3:* Calculate wired budget as E-$W_B$
    *step 4:* Give ONU minimum bandwidth based on number of clients and client types supported, call this $W_F$
    *step 5*: Excess bandwidth is calculated as E – $W_F$ – $W_B$
    *step 6:* Allocate excess bandwidth in proportion to average call drop for the ONU (as proposed in [18])

**Algorithm 5: Highest-Priority-First**
**Input:** Data rates for each BS, total EPON bandwidth E
**Output:** Bandwidth assignment for each ONU, MoS for wired clients
    *step 1:* Allocate backhaul resources to all wireless BS based on wireless resource allocation
    *step 2:* Calculate allocated budget in step 1, call it $W_B$
    *step 3:* Calculate wired budget as E-$W_B$
    *step 4:* Give each ONU minimum bandwidth based on number of clients and types supported, call this $W_F$
    *step 5: Excess bandwidth is calculated as $E - W_F - W_B$*
    *step 6: Allocate excess bandwidth to highest-priority clients so that they achieve maximum possible MoS, and then move on to lower priority till excess bandwidth > 0*

speed. Figure 4 plots cumulative call-drop probability due to user satisfaction using our algorithms at high, medium, and low loads and confirms the observations seen for MoS..

From Figs. 5-6, we can see that Google+ provides similar results to Skype [25]. More precisely, Google+ has slightly higher MoS and lower call-drop compared to Skype (Figs. 3-4) due to higher number of resolutions supported and hence making solution space larger for QoE optimization.

MoS and call-drop probability, averaged for each client type, are shown in Figs. 7 and 8 for low, medium, and high-load scenarios. We see that WF performs slightly better than MRR for each client type. However, since these algorithms do not exploit the characteristics of the application like MCKP, MCKP gives the best QoE for each client type at medium to high loads. The performance gain achieved by MCKP increases as the load increases where there is higher contention for bandwidth.

*A. Evaluation of EPON Resource-Allocation Schemes*

MoS of application-aware EARA are compared with application-unaware HPF in Fig. 9 for different client types at low, medium, and high loads. EARA outperforms HPF in providing higher MoS for client types 1 and 2 (lower client types) for all load scenarios. EARA uses application information to reduce user unsatisfaction (call-drop probability). Knowledge of the application makes the controller take decisions that would improve overall QoE of the system by giving appropriate amounts of bandwidths to each client type. For applications like Skype, where there is no linear relation between user satisfaction and bandwidth provisioning, there needs to be application awareness to help decide on resource allocation. HPF strives to assign maximum achievable MoS (and hence bandwidth) to higher-priority customers without weighing the loss incurred to the lower-priority clients. So, we see that, at high load, client type 1 suffers largely from low MoS using HPF, but EARA balances between different client types after providing minimum acceptable MoS to each client.

## VI. CONCLUSION

This study demonstrates that resource-allocation schemes in converged OFDMA wireless networks and EPON that utilize application information can outperform application-unaware schemes in terms of Quality of Experience (QoE) measures such as empirical Mean Opinion Score (MoS). The proposed architecture provides better client-and-service-level differentiation by considering different QoE requirements in terms of minimum MoS for each client type. The proposed resource-allocation schemes in both wired and wireless networks strive to maximize MoS of the system after providing minimum MoS to each client.

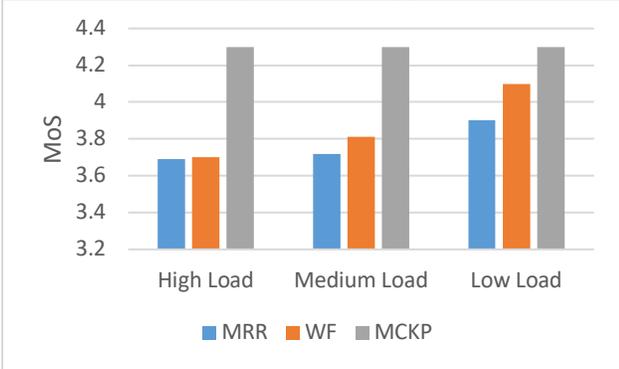

Figure 3: Cumulative MoS for Skype profile wireless resource allocation.

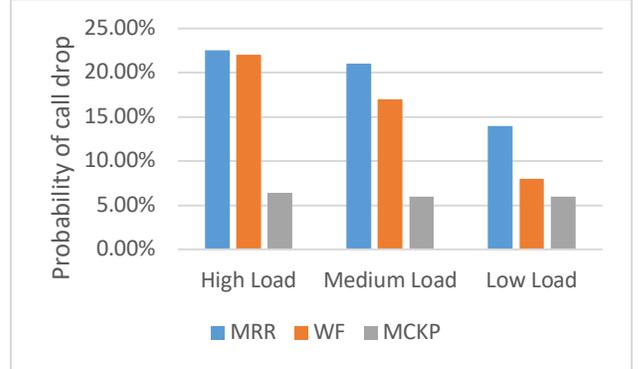

Figure 4: Cumulative call-drop probability for Skype profile wireless resource allocation.

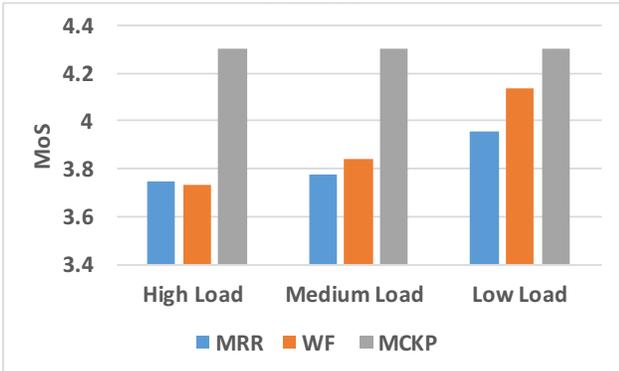

Figure 5: Cumulative MoS for Google+ profile wireless resource allocation.

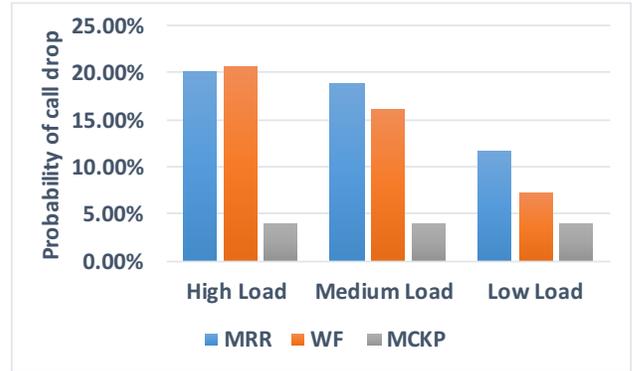

Figure 6: Probability of call drop for Google+ profile wireless resource allocation.

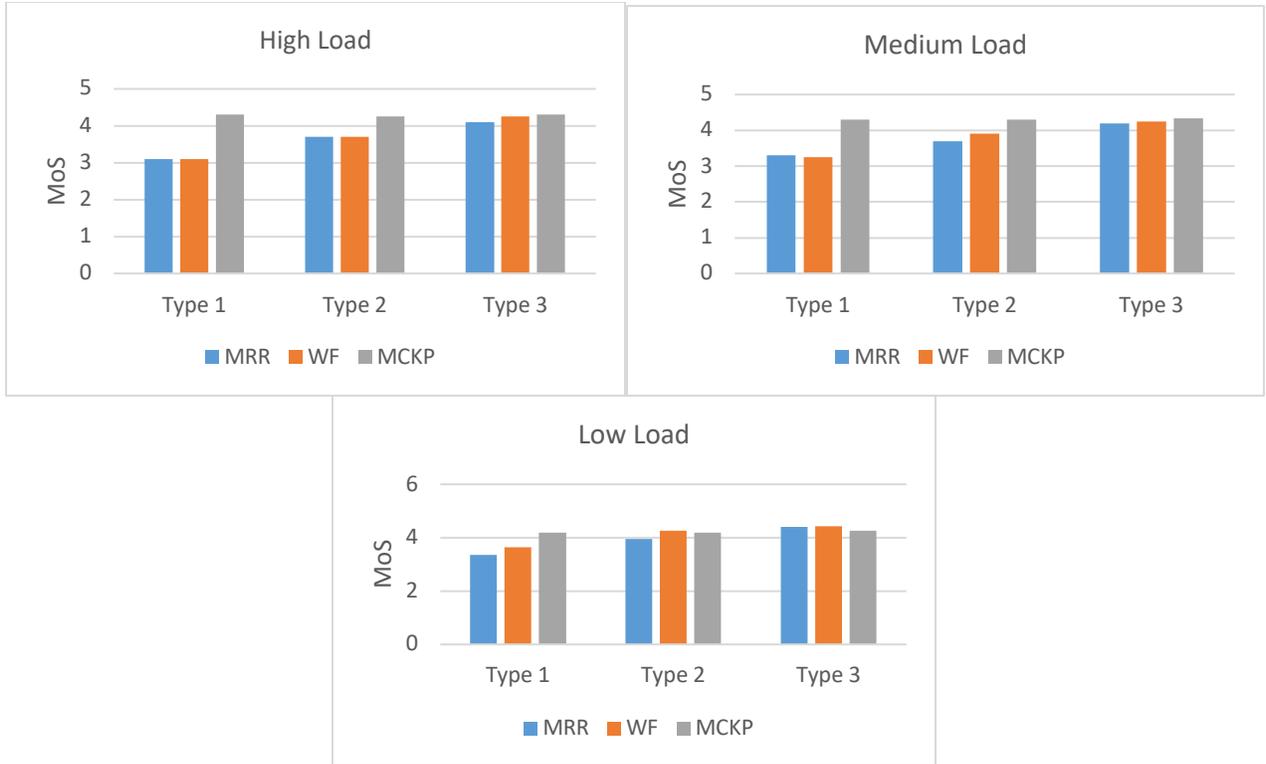

Figure 7: Average MoS for each client type for wireless resource-allocation algorithms for high, medium, and low-load scenarios.

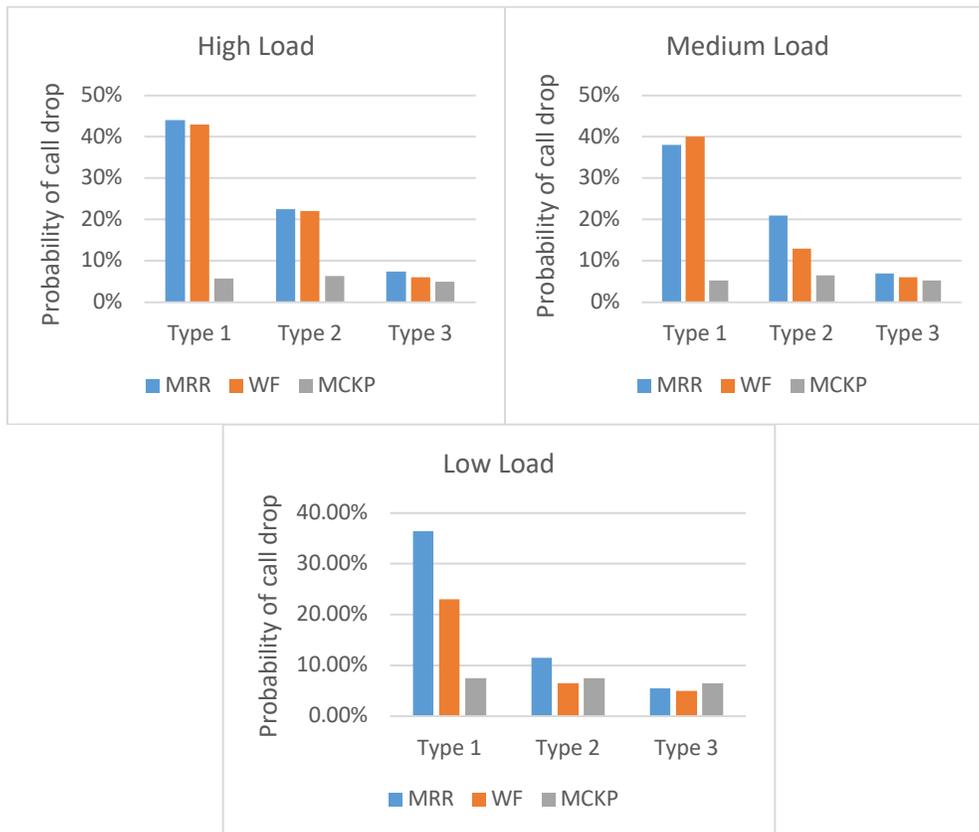

Figure 8: Average user satisfaction (call-drop probability) for each client type for wireless resource-allocation algorithms for high, medium, and low-load scenarios.

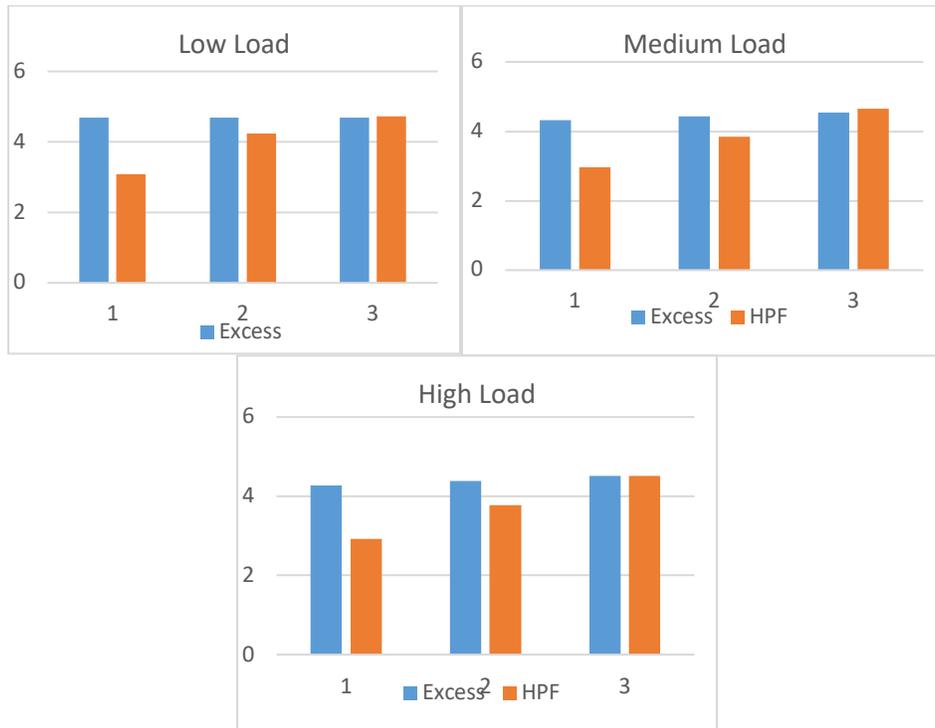

Figure 9: Average MoS at each client type for wired resource allocation at low, medium, and high loads.

## ACKNOWLEDGMENT

This work was supported by the Republic of Korea's ICT R&D program of MSIP/IITP [B0132-16-1004, SDN based wired and wireless converged optical access networking].

...